\documentclass[journal]{IEEEtran}

\ifCLASSINFOpdf
\else
   \usepackage[dvips]{graphicx}
\fi
\usepackage{url}

\usepackage{float}
\usepackage{graphicx}
\usepackage{tikz}
\usetikzlibrary{math, shapes}
\usepackage{pgfplots}
\pgfplotsset{
    compat=1.9, 
    legend style={font=\footnotesize, fill opacity=0.6, draw opacity=1, text opacity=1, draw=white!15!black, legend cell align=left, align=left}, 
    width=7.5cm, 
    grid style={line width=.1pt, draw=gray!10},
    major grid style={line width=.2pt,draw=gray!50},}
\usepackage{subcaption}
\usepackage[normalem]{ulem}
\usepackage{multirow}
\usepackage{array}
\usepackage{booktabs}
\usepackage[T1]{fontenc}
\usepackage[nospace,noadjust]{cite}
\usepackage[nolist,printonlyused]{acronym}
\usepackage{comment}
\usepackage{mleftright,xparse}
\usepackage{amssymb,amsmath,amsfonts,amsthm,bm,mathtools,cuted}
\usepackage[ruled,vlined,linesnumbered]{algorithm2e}

\newcommand{\E}{\mathbb{E}}
\newcommand{\set}[1]{\left\{#1\right\}}
\newcommand{\mc}[1]{\mathcal{#1}}
\newcommand{\mb}[1]{\mathbf{#1}}

\DeclareMathOperator*{\argmax}{arg\,max}

\newcommand{\tc}{\,|\,}

\newcommand{\FS}[1]{\textcolor{red}{{#1}}}
\newcommand{\edit}[1]{\textcolor{black}{#1}}
\newcommand{\reedit}[1]{\textcolor{black}{#1}}

\theoremstyle{plain}
\newtheorem{thm}{Theorem}

\begin{document}

\title{\edit{WMMSE resource allocation for  FD-NOMA}}
\author{Andrea Abrardo, Marco Moretti and Fabio Saggese
\thanks{Andrea Abrardo (abrardo@dii.unisi.it) is with the DIISM, University of Siena, Italy and CNIT; Marco Moretti (marco.moretti@unipi.it) is with the DII, University of Pisa, Italy and CNIT; his work is partially supported by the MIUR in the CrossLab project; Fabio Saggese (fasa@es.aau.dk) is with ES Department, Aalborg University, Denmark.  Corresponding author: F. Saggese.}}
\maketitle

\begin{abstract}
\reedit{Resource} allocation in interference-limited systems is a key enabler for beyond 5G (B5G) technologies, such as multi-carrier full duplex non-orthogonal multiple access (FD-NOMA). In FD-NOMA systems \reedit{resource} allocation is a  \reedit{computation-intensive} non-convex problem due to the presence of strong interference and the integrality condition on channel allocation. In this paper, we propose \reedit{an iterative algorithm based on the combination of channel and power allocations} aimed at the minimization of the weighted mean square error, which converges to a feasible allocation of the original problem. \edit{Experimental results show that the proposed algorithm  \reedit{has lower complexity than} other state-of-the-art solutions for the same problem. Moreover,} \reedit{the presented results}
 assess the validity of our approach showing performance close to the theoretical optimum.
\end{abstract}

\IEEEpeerreviewmaketitle

\section{Introduction} \label{Sec:Intro}

\IEEEPARstart{F}{ull} duplex (FD) and non-orthogonal multiple access (NOMA) are among the most promising technologies to be adopted for future wireless communication systems~\cite{Elbayoumi2020}.
These two technologies achieve superior performance by relaxing the conventional requirement of orthogonal access to the wireless channel. NOMA and FD have been recently combined in a novel scheme, which has the advantage of achieving enhanced flexibility, user fairness, and increased spectral efficiency \cite{Mohammadi2019}. The price to pay is that the system performance is severely affected by the presence of large multi-user interference so that power and channel  allocation~\cite{Ding2016, DellaPenda2019} are fundamental for its successful implementation and deployment. 
As a matter of fact, even in a single carrier scenario, power allocation in FD-NOMA is not convex due to the mutual interference among users. Hence, its solution requires the use of advanced and complex algorithms, e.g., see~\cite{Nhat2021}. In a multi-carrier scenario, the complexity increases since one has to jointly establish the allocation and the pairing order on each subcarrier. 
To the best of our knowledge, only a few works in the literature have addressed the problem of resource allocation for FD-NOMA systems~\cite{Ding2018a, Sun2017, Nguyen2019a, Abrardo2020}. All these studies \edit{consider the case where both the base station and the nodes have a single antenna. On the other hand, in the MIMO case users can be separated by a beamforming precoder, and the NOMA paradigm  can be applied to users belonging to the same beam ~\cite{Kimy2013, Choi2015, Chen2016, Chen2017,Saggese2020}.} The FD-NOMA solutions proposed so far envisage optimization algorithms having large to very large computational loads. The main reason for their complexity is that: a) the objective function is not convex; b) the allocation problem involves a set of mixed allocation variables: \emph{binary}, the channel assignment, and \emph{continuous}, the power allocation.
In~\cite{Sun2017, Nguyen2019a}  the binary assignment variables have been addressed by first relaxing the integrality condition and then assigning a penalty method for non-integer variables. Using an alternative perspective, in~\cite{Abrardo2020} we have proposed a scheme that solves the problem in the dual Lagrangian domain, following a block coordinate descent approach. 

In this work, we propose a \emph{low-complexity} resource allocation algorithm for the rate maximization in multi-carrier \edit{SISO} FD-NOMA systems. \reedit{Rather than jointly optimizing channel assignment and power allocation, the proposed algorithm follows a suboptimal layered approach based on two steps:}
\vspace{-0.5cm}\reedit{\begin{enumerate}
    \item \emph{Channel allocation}: Allocate  in a FD fashion every channel to one user in each transmit direction; 
   \item \emph{Power allocation}: Distribute the power with the goal of minimizing of the weighted mean square error (MSE)~\cite{Shi11}.
\end{enumerate}
One of the main contributions of this letter is to analytically prove  that for the additional users  employing the NOMA paradigm \emph{channel allocation is an implicit result of power allocation}: each resource is assigned to the user having a non-zero power coefficient on that channel. Accordingly,  after step 1) it is possible to \emph{relax} the channel allocation requirements and just  focus mainly on power allocation, thus noticeably reducing the problem complexity.}
Simulations show results close to the  optimum, proving the effectiveness of our approach.

\section{System Model} \label{sec:model}
We consider an OFDMA FD-NOMA system, where single-antenna user equipments (UEs) are served by a single-antenna base station (BS). Due to hardware limitations, the FD technology is implemented only at the BS, which is able to cancel a large fraction of the self-interference that it generates.
The NOMA paradigm is implemented at both the BS and the UEs, which are able to cancel a certain number of interfering signal of the same type (uplink or downlink) on each channel $f\in \mc{F}$, the set of available channels, through successive interference cancellation (SIC). To correctly perform SIC, we assume that a) the signal to be cancelled is perfectly reconstructed, condition assumed for the remainder of the paper, and b) absence of detection errors, condition discussed in Section~\ref{sec:canc}.

We denote by $\mathcal{U}$ and $\mathcal{D}$ the sets of the $M = \left|\mathcal{U}\right|$ uplink and  $N = \left|\mathcal{D}\right|$ downlink users in the system, respectively. 
The information symbols $s_{i,f}$ transmitted (or received) by user $i$ on subcarrier $f$ are modelled as zero-mean complex random variables with with unitary power,
the  signal  is then scaled  by a factor $\sqrt{P_{i,f}}$, so that the signal power is $P_{i,f}$\edit{, where the index $i$ denotes the transmitter if $i\in\mathcal{U}$ or the receiver if $i\in\mathcal{D}$}.
In the FD-NOMA scenario all users can transmit on any subcarrier without any orthogonality requirements and uplink and downlink transmissions are both affected by  uplink and downlink interference. \edit{This is a particularly challenging scenario since uplink users might cause serious interference to downlink users if proper countermeasures are not in place. In this setting,} the signal intended for user $i \in \mc{U}\cup \mc{D}$ is 
\begin{equation} \label{eq:genericSignal}
\begin{aligned}
{y}_{i,f} = h_{i,i}(f)\sqrt{P_{i,f}}{s}_{i,f}+\hspace{-0.4cm}\sum\limits_{j \in \mc{U}\cup\mc{D}\setminus i}\hspace{-0.3cm}h_{j,i}(f)\sqrt{P_{j,f}}{s}_{j,f} + z_{i,f},
\end{aligned}
\end{equation}
where 
$z_{i,f}$ is the zero-mean thermal noise at the receiver with variance $\sigma^{2}=\E\left\{|z_{i,f}|^2\right\}$, $h_{i,i}$ represents the direct channel between UE and the BS and $h_{j,i}$ describes  \edit{the multi-user (MUI)   due to NOMA and co-channel interference (CCI) due to FD. Accordingly, depending on the combination of uplink and downlink users there
are four different cases for the interference} a) if $i \in \mc{D}$ and $j \in \mc{U}$, $h_{j,i}(f)$ represents the cross-channel gains between uplink and downlink users \edit{(CCI)}; b) if $i \in \mc{U}$ and $j \in \mc{D}$, $h_{j,i}(f)$ denotes the residual gain due to non perfect self-interference cancellation at the BS \edit{(CCI)}; c) if $i,j\in\mathcal{U}$, it is $h_{j,i}(f)=h_{j,j}(f)$ \edit{(MUI)} and d) if $i,j\in\mathcal{D}$, it is $h_{j,i}(f)=h_{i,i}(f)$ \edit{(MUI)}.    

\subsection{Interference cancellation}\label{sec:canc}
For practical reasons, we assume that for any subcarrier the message of  \emph{at most} two users can be sent in any direction and of those two users, according to the NOMA paradigm, only one is able to \emph{cancel interference} coming from the the same direction. For both uplink and downlink, we single out a \emph{strong} and a \emph{weak user}. Their definition is asymmetric: in the downlink,  strong users  cancel the contributions of the other users, while in the uplink the contribution of the strong user can be easily canceled by  the BS so that the weak users are not interfered by it. Accordingly, we introduce the binary allocation variable $x_{i,f}\in\mc{X}$, which is set to 1 if user $i$ is the strong one on channel $f$ and 0 otherwise. The set $\mc{X}$ is defined consistently with our assumptions as 
$\mc{X}=\left\{x_{i,f}\in\{0,1\}\,|\,i\in\mc{U}\cup\mc{D},\sum_{i\in\mc{U}} x_{i,f}\le 1,\,\sum_{i\in\mc{D}} x_{i,f}\le 1\right\}$.
After  SIC the received signal on subcarrier $f$ for user $i$ is
\begin{equation}
{y}_{i,f}= h_{i,i}(f)\sqrt{P_{i,f}}{s}_{i,f} + \sum_{j \in \mathcal{I}_{i,f}^{(x)}}h_{j,i}(f)\sqrt{P_{j,f}}{s}_{j,f}+ z_{i,f},
\label{eq: fusion}
\end{equation}
where $\mc{I}_{i,f}^{(x)}$ is the set of the  \emph{potential} users interfering with user $i$ on subcarrier $f$, and it is defined as
\begin{equation} \label{eq:interferenceSet}
\mc{I}_{i,f}^{(x)}=
\begin{cases}
\mc{U} \cup \mc{D} \setminus \set{i,s} &i,s \in \mc{U} , x_{s,f} = 1, \\
\mc{U} &i \in \mc{D}, x_{i,f} = 1, \\
\mc{U} \cup \mc{D} \setminus i &i \in \mc{D}, x_{i,f} = 0 \parallel i \in \mc{U}, x_{i,f} = 1.
\end{cases}
\end{equation} 
In practice, the number of actively interfering users on channel $f$ is smaller than the cardinality of $\mc{I}_{i,f}^{(x)}$. For example, user $k$, which sets $P_{k,f} =0$, does not interfere with $i$ on channel $f$ even if it belongs to $\mc{I}_{i,f}^{(x)}$. The advantage of the formulation~\eqref{eq: fusion} is that does not require an explicit allocation of the weak user but depends only on power allocation and as such is helpful for the derivation of the main algorithm in Section~\ref{sec: powAl}.

Let us denote by $\mb{P}_{\mathcal{D}} = \left\{P_{i,f}\right\}$, $f \in \mc{F}$, $i \in \mc{D}$ and $\mb{P}_{\mathcal{U}} = \left\{P_{i,f}\right\}$, $f \in \mc{F}$, $i \in \mc{U}$ the vectors collecting the transmit powers for all downlink and uplink users in the system and by $\mb{P} = [\mb{P}_{\mathcal{D}} ,\mb{P}_{\mathcal{U}}]$,   
from \eqref{eq: fusion}, the signal to interference-plus-noise ratio (SINR) for user $i$ on subcarrier $f$ is 
\begin{equation}
\begin{aligned}
&\gamma_{i,f}(\mb{P},x) = \frac{\left|h_{i,i}(f)\right|^2 P_{i,f}}
{\sum_{j \in  \mc{I}_{i,f}^{(x)}} \left|h_{j,i}(f)\right|^2 P_{j,f} + \sigma^2 }  ,
\end{aligned}
\label{eq: SINR}
\end{equation}
so that the rate of $i$ on $f$ is $R_{i,f}(\mb{P},x)= \log\left(1+\gamma_{i,f}(\mb{P},x)\right)$.

In the uplink, interference cancellation is straightforward: the  BS receives all the data streams and, hence, can always  successfully cancel the strong user. Conversely, in the downlink  the  data stream intended  for the weak user's receiver is canceled at the strong user's receiver. Thus, if $s\in\mc{D}$ is the strong user and $k\in\mc{D}$ is the weak one  on the downlink channel $f$, the condition for perfect cancellation of $k$  is
\begin{equation}
R_{k,s,f}(\mb{P},x)>R_{k,f}(\mb{P},x)
\end{equation}
where $R_{k,s,f}(\mb{P},x)$ is the achievable rate of user $k$ measured at the receiver $s$.
This condition is equivalent to~\cite{Sun2017}
\begin{equation}
\label{eq:GAMMA}
\Gamma_{k,s}(f) = \textstyle\sum_{j \in \mc{U}}\Theta_{j,f}^{(k,s)}P_{j,f}
+\Delta_{f}^{(k,s)}\leq 0,
\end{equation}
with $\Theta_{j,f}^{(k,s)} =|h_{k,k}(f), h_{j,s}(f)|^2- |h_{s,s}(f), h_{j,k}(f)|^2 $ and $\Delta_{f}^{(k,s)}= \left(|h_{k,k}(f)|^2 - |h_{s,s}(f)|^2 \right) \sigma^2 $. 

\section{Max sum-rate algorithm for FD-NOMA} \label{sec:wmmsemodel}
To formulate the max-rate optimization problem, we introduce also the binary allocation variable $t_{i,f}\in\mc{T}$, which is set to 1 if user $i$ is the weak one on channel $f$ and 0 otherwise. Taking into account that there are at most two users per channel direction and that a user can not be at the same time weak and strong on a given subcarrier, 
$\mc{T}$ is defined as the set $\mc{T}=\{ t_{i,f} \in \{ 0,1 \} \,|\, i\in\mc{U}\cup\mc{D},\,t_{i,f}=0 \text{ if } x_{i,f}=1,\,\sum_{i\in\mc{U}} t_{i,f}\le 1,$ $\sum_{i\in\mc{D}} t_{i,f}\le 1\}$.
Our objective  is to allocate power and channels to the users to maximize the overall weighted sum-rate $U(x,t,\mb{P}) =\sum_{f\in\mathcal{F} } \sum_{i \in \mathcal{U}\cup\mathcal{D}} x_{i,f}\alpha_s R_{i,f}(\mb{P},x)+t_{i,f}\alpha_{w} R_{i,f}(\mb{P},x)$
\begin{align} 
\label{Rmax}
    &\max_{\substack{\mathbf{P} \succeq 0, \\x\in\mc{X},\,t\in\mc{T}}} U(x,t,\mb{P})  \\
    & \hspace{3 cm}\text{subject to} \notag \\ 
        &\hspace{.5 cm}\sum\limits_{f \in \mc{F}} P_{i,f} \le P_{U}, \quad  \forall i \in \mathcal{U} \tag{\ref{Rmax}.a} \label{WMMSE:PU}  \nonumber\\
        & \hspace{.5 cm}\sum\limits_{i \in \mathcal{D}}\sum\limits_{f \in \mc{F}} {P}_{i,f} \le P_{D}  \tag{\ref{Rmax}.b}  \label{WMMSE:PD}  \nonumber\\        
        &\Gamma_{k,s}(f) \leq 0, \begin{cases} k,s \in \mc{D} \tc x_{s,f} = 1,t_{k,f} = 1,\\
       P_{k,f}>0,P_{s,f}>0,  \end{cases} \hspace{-0.2cm} \forall f\in \mc{F}. \tag{\ref{Rmax}.c}\label{WMMSE:WSWW}     
\end{align}
The positive weights $\alpha_{s}$ and $\alpha_{w}$ in the objective function are employed to enforce a certain degree of fairness among strong ad weak users, constraints~\eqref{WMMSE:PU} and~\eqref{WMMSE:PD} are the power budget for the uplink and downlink users and constraint~\eqref{WMMSE:WSWW} guarantees successful pairwise SIC at downlink strong user $s$ when the weak user $k$ is actually transmitting on the same subcarrier. Unfortunately,~\eqref{Rmax} is a mixed integer programming problem because of the simultaneous presence of binary and continuous variables, and this, together with the 
interference cancellation constraint, makes it extremely complex to solve. Accordingly, we choose a heuristic approach to~\eqref{Rmax} breaking the solution of the optimization in three phases:
\begin{enumerate}
    \item \emph{Allocate} the strong users: in this phase one strong uplink and one strong downlink user is selected for each subcarrier.
    \item \reedit{Fixing the allocation of phase 1)}, \emph{relax} the constraint that only one weak user is allowed for channel: the number of weak users active on a channel is dictated by the number of users whose power is non-zero \reedit{on that channel}. Channel assignment becomes the result of power allocation. Anyhow, each weak user still needs to satisfy~\eqref{eq:GAMMA} to be \emph{feasible}. 
    \item \emph{Power allocation}: having fixed $x\in\mc{X}$ and relaxed $t$, \eqref{Rmax} becomes the power allocation problem
\end{enumerate}
\vspace{-0.05cm}
\begin{align} \label{Rmax2}
    \max_{\mathbf{P} \succeq 0}&\sum_{f \in \mc{F}} \hspace{-1pt} \sum_{i \in  \mathcal{U}\cup\mathcal{D}} \hspace{-2pt}     \big(x_{i,f}\alpha_{s} + (1-x_{i,f})\alpha_{w}\big) R_{i,f}(\mb{P},x)  \\
    & \text{subject to}~\eqref{WMMSE:PU},~\eqref{WMMSE:PD}, ~\eqref{WMMSE:WSWW}.\notag
\end{align}
\edit{The only part where binary assignment is needed is in the first step of the algorithm, i.e., the  strong users allocation part, which is a pure full-duplex allocation problem solvable with state-of-the-art approaches such as~\cite{abrardo2018optimal}.}

The fact that more than one weak user is allowed per subcarrier simplifies the formulation of problem~\eqref{Rmax2} but could potentially raise many practical problems. 
Nevertheless, we can formulate the following theorem.
\begin{thm} \label{theo:twouser}
Any local optima for~\eqref{Rmax2} allows \reedit{the allocation of} at most two NOMA users per channel in each direction.
\end{thm}
\begin{proof}
See Appendix.
\end{proof}
\noindent 
\edit{Thus, any local optimum solution of the relaxed problem~\eqref{Rmax2} is compliant with the exclusive integer constraints on channel assignment \reedit{of problem~\eqref{Rmax}, i.e., feasible channel assignment is a results of power allocation.}}

\subsection{A WMMSE formulation for \eqref{Rmax2}} \label{sec: powAl}
Even after removing the dependence on the binary allocation variables $x$ and $t$, the power allocation  \eqref{Rmax2} is \emph{not convex}: in the sum-rate expression the power of a given user $k$ appears both at the numerator of the SINR when $k$ is the desired user and at the denominator when $k$ is seen as interference. To address the non-convexity, we reformulate the sum-rate maximization problem in the presence of interference  as weighted MSE minimization \cite{Shi11}. Regardless of the transmit direction, multiplying the received signal $y_{i,f}$ by a scaling factor $g_{i,f}$ yields the MSE  
\begin{equation}
\begin{aligned}
{e}_{i,f} = \, &\mathbb{E}\left\{\left|{g}_{i,f} y_{i,f}-s_{i,f}\right|^2\right\} =|1-{g}_{i,f}h_{i,i}(f) \sqrt{{P}_{i,f}}|^2 \\ & \quad + \sum\limits_{\mathclap{j \in  \mathcal{I}_{i,f}^{(x)}}}\left|{g}_{i,f}h_{j,i}(f)\right|^2 P_{j,f} +  |g_{i,f}|^2\sigma^2.
 \end{aligned}
\label{eqmse}
\end{equation}
By differentiating~\eqref{eqmse} and setting the derivative to zero, the value of  $g_{i,f}$ that minimizes the MSE is
\begin{equation}
g_{i,f} = \frac{{h}_{i,i}^*(f) \sqrt{P_{i,f}}}
{\left|h_{i,i}(f)\right|^2P_{i,f} + \sum_{j \in  \mc{I}_{i,f}^{(x)}} \left|h_{j,i}(f)\right|^2P_{j,f} + \sigma^2},
\label{mmsefilterscalar}
\end{equation}
and the correspondent value for the minimum MSE is
\begin{equation}
\label{eq:mmseRel}
e_{i,f}(\mb{P}) = (1+\gamma_{i,f}(\mb{P},x))^{-1}.
\end{equation} 

The weighted minimum MSE problem is defined as 
\begin{align} \label{WMMSE:RA3}  
\min_{ \mathbf{P}\succeq 0, \mathbf{w}\succeq 0, \mathbf{g}} & \sum\limits_{f \in \mc{F}} \sum\limits_{i \in \mathcal{U}\cup\mathcal{D}}  \alpha_{i,f}  \left[w_{i,f}e_{i,f}(\mb{P})-\log(w_{i,f})\right]\\
& \text{subject to}~\eqref{WMMSE:PU},~\eqref{WMMSE:PD}, ~\eqref{WMMSE:WSWW}.\notag
\end{align}
where $\alpha_{i,f} = x_{i,f} \alpha_s + (1 - x_{i,f}) \alpha_w$ is $\alpha_s$ if user $i$ is the strong one on channel $f$ or $\alpha_w$ otherwise, $w_{i,f}$ are positive weights. $\mathbf{w}$ and $\mathbf{g}$ are the vectors collecting all values of $w_{i,f}$ and $g_{i,f}$, respectively. We can now formulate the theorem that allows us to solve~\eqref{Rmax2} as a weighted MSE minimization.
%
\begin{thm} \label{theo:equiv} Solving~\eqref{WMMSE:RA3} is equivalent to solving~\eqref{Rmax2}, i.e.  \eqref{WMMSE:RA3} yields a local optimum for~\eqref{Rmax2}.
\end{thm}
\begin{proof}
See~\cite{Shi11} for the equivalence of the problems of rate maximization and weighted MSE minimization; \edit{a specific proof for~\eqref{Rmax2}-\eqref{WMMSE:RA3} is in~\cite{arxProof}, omitted here for space limits.}
\end{proof}

Problem~\eqref{WMMSE:RA3} is still not convex, but can be solved by adopting  an iterative \emph{block coordinate descent} (BCD) scheme.
BCD converges to a local minimum if the optimization over each block of variables is convex and differentiable \cite{Tseng2001}. Thus, we split the original problem \eqref{WMMSE:RA3} into four different sub-problems obtained by fixing any three out of the four sets $\mathbf{P}_{\mathcal{D}}, \mathbf{P}_{\mathcal{U}}, \mathbf{w}$ and $ \mathbf{g}$  of optimization variables and  solve~\eqref{WMMSE:RA3} with respect to (w.r.t.) the remaining set of variables. By doing so, \emph{each of the sub-problems is convex} and differentiable and, iterating the solution of the problems sequentially, the algorithm converges to a local optimum of~\eqref{WMMSE:RA3}, which, because of Theorem~\ref{theo:equiv},  yields a power distribution that locally maximizes~\eqref{Rmax2} as well. 
\paragraph{\textbf{Optimizing w.r.t. $\mathbf{g}$.}} Given the power allocations $\mathbf{P}$, $g_{i,f}$ $\forall i\in \mc{U}\cup\mc{D}$, $f\in\mc{F}$ can be evaluated as in~\eqref{mmsefilterscalar}.

\paragraph{\textbf{Optimizing w.r.t. $\mathbf{w}$.}} Given $\mathbf{P}$ and $\mathbf{g}$, we can compute $e_{i,f}$ as in~\eqref{eqmse}. Differentiating the objective function w.r.t. $\mathbf{w}$ and setting the result to zero yields
\begin{equation}
\label{eq:w}
{w}_{i,f} = e_{i,f}^{-1} \quad  \forall i\in \mc{U} \cup\mc{D}, \forall f \in \mc{F}.
\end{equation}
Since $0<e_{i,f}\leq 1$, the condition $\mathbf{w} \succeq 0$ is always met.

\paragraph{\textbf{Optimizing w.r.t. $\mathbf{P}_{\mathcal{D}}$.}}
Given the values of $\mathbf{g}$, $\mathbf{w}$ and $\mathbf{P}_{\mathcal{U}}$,  the power allocation problem for the downlink users is convex and can be solved in the dual domain. Neglecting irrelevant terms, the Lagrangian dual function is
\begin{equation}
\begin{aligned}
\min_{ \mathbf{P}_{\mathcal{D}} \succeq 0 } &
\sum\limits_{f \in \mc{F}} \sum\limits_{k \in \mathcal{D}_{f} } \beta_{k,f} e_{k,f}(\mb{P}) + \mu P_{k,f},
\end{aligned}
\label{eq:dualDown}
\end{equation} 
where $\beta_{k,f} = \alpha_{k,f} w_{k,f}$ and $\mu$ is the positive Lagrangian multiplier associated to the constraint~\eqref{WMMSE:PD}. Constraint~\eqref{WMMSE:WSWW} is 
translated in an on/off condition: given channel $f$ and the strong downlink user $s$, a weak user $k$ can transmit only if $\Gamma_{k,s}(f) \le 0$. This is captured in~\eqref{eq:dualDown} by the set $\mathcal{D}_{f} = \mathcal{D}^w_{f} \cup \{s\,|\,x_{s,f}=1\}$, which includes the strong downlink user $s$ and $\mathcal{D}^w_{f}=\{k\,|\,x_{k,f}=0\,\&\,\Gamma_{k,s}(f)\le0\}$, the set of the weak users satisfying~\eqref{WMMSE:WSWW}.
The optimal power for $i \in \mc{D}_{f}$ is
\begin{align} 
{P}_{i,f} = \hspace{-3pt} \frac{ \left|\beta_{i,f}g_{i,f} h_{i,i}(f)\right|^{2}}{\Big(\beta_{i,f} \left|{g}_{i,f}h_{i,i}(f)\right|^2 \hspace{-4pt}+\hspace{-8pt} \sum\limits_{\ell \in   \mathcal{C}_{i,f}^{(x)}} \hspace{-8pt} \beta_{\ell,f} \left|{g}_{\ell,f}h_ {i,\ell}(f)\right|^2 \hspace{-4pt}+\hspace{-3pt} \mu \Big)^2}
\label{eq:powDown}
\end{align}
and it is zero for all the other  users. 
In~\eqref{eq:powDown}, we have used the set $\mathcal{C}_{i,f}^{(x)}$, which  represents the set of users that potentially receive interference from $i\in\mathcal{U}\cup\mathcal{D}$ on subcarrier $f$, i.e. 
\begin{equation} \label{eq:Cinterference}
\mc{C}_{i,f}^{(x)} =
\begin{cases}
\mc{D}, &i \in \mc{U}, x_{i,f} = 1, \\
\mc{U} \cup \mc{D} \setminus i, &i \in \mc{U}, x_{i,f} = 0 \parallel i \in \mc{D}, x_{i,f} = 1,\\
\mc{U} \cup \mc{D} \setminus \set{i,s}, &i,s \in \mc{D}, x_{s,f} = 1.
\end{cases}
\end{equation}

\paragraph{\textbf{Optimizing w.r.t. $\mathbf{P}_{\mathcal{U}}$.}}
Given the values of $\mb{g}, \mb{w}$ and $\mb{P}_{\mathcal{D}}$, the problem evaluating $\mb{P}_{\mc{U}}$ 
is convex. The extra NOMA condition~\eqref{WMMSE:WSWW}, which is active only for the uplink users, is linear in $\mathbf{P}_{\mathcal{U}}$, and the Lagrangian   dual function is
\begin{equation}
\min_{ \mathbf{P}_{\mathcal{U}} \succeq 0 }
\sum\limits_{j \in \mathcal{U} } \sum\limits_{f \in \mc{F}} \beta_{j,f}e_{j,f} +\mu_{j} P_{j,f} +
\sum\limits_{k \in \mathcal{D}^w_f } \mu_{k,f} \Theta_{j,f}^{(k,s)}P_{j,f}
\label{eq:dualUp}
\end{equation}
where $\mu_j$ and $\mu_{k,f}$ are positive Lagrangian multipliers associated to constraint~\eqref{WMMSE:PU} and~\eqref{WMMSE:WSWW}, respectively and $s$ is the strong downlink user on $f$, i.e., $s \in \mc{D} \tc x_{s,f} = 1$.
The power coefficient for user $i \in \mc{U}$ can be evaluated as
\begin{equation}
{P}_{i,f} = \hspace{-3pt} \frac{\left|w_{i,f} g_{i,f} h_{i,i}(f) \right|^2}{\Big(|{g}^*_{i,f}h_{i,i}(f)|^2 +\hspace{-4pt} \sum\limits_{\ell \in   \mathcal{C}_{i,f}^{(x)}} \hspace{-4pt} w^*_{\ell,f} |{g}^*_{\ell,f}h_ {i,\ell}(f)|^2 \hspace{-3pt} + \hspace{-3pt}\lambda_i \Big)^{2}}
\label{eq:powUp}
\end{equation}
where,  $\lambda_i = \mu_i + \sum_{k \in \mathcal{D}} \mu_{k,f} \Theta_{i,f}^{(k,s)}$, and $\mathcal{C}_{i,f}(x)$ is given in~\eqref{eq:Cinterference}. 
The evaluation of $\mu_i$ and $\mu_{k,f}$ is obtained through well known ellipsoid method~\cite{Yu2006}.
\section{Numerical Results}
\label{Sec:num_res}
We consider a single cell scenario with $F = 6$ subcarriers, and with the same number of uplink and downlink users $M = N$. \edit{The number of channels and users is set to test the system in overloaded conditions~\cite{Abrardo2020}.}  The cell radius is 100 m\edit{~\cite{Nguyen2019a}}. 
The path loss exponent is $4$, while the shadowing is log-normally distributed having standard deviation $8$ dB. 
The SI cancellation factor at the BS is set to a constant value of 110 dB, as in~\cite{Sun2017}.
\edit{The strong users needed to initialize the WMMSE algorithm are selected by maximizing the overall sum-rate neglecting the interference due to NOMA paradigm. This problem is solved running the full-duplex allocation algorithm given in~\cite{abrardo2018optimal}.} The value of the weights for the strong users is set to $\alpha_s = 1$, while the value of $\alpha_w$ varies depending on the simulations. For all the results, the maximum power for each uplink user is $P_U = 14$ dBm and $P_D = 20$ dBm for the BS.
\reedit{As benchmark, we use} the theoretic optimum (REF) and the sub-optimal (SCA) approach proposed in~\cite{Sun2017}.


\edit{Fig.~\ref{fig:convergence} shows the performance of the proposed and benchmark algorithms in terms of the weighted sum-rate $U$, for different numbers of users in the cell, and with $\alpha_w = 2$.}
\edit{We plot the performance of the proposed approach (WMMSE), and of the pure full-duplex approach (FD-OMA)~\cite{abrardo2018optimal}, as a function of the number of iterations. We also show the performances attained at the converge by REF and SCA benchmark algorithms~\cite{Sun2017}.}
\edit{For WMMSE, $U$} increases monotonically after each iteration, as proven in~\cite{arxProof}. 
Since after the firsts 100 iterations the increment of performances is negligible, \edit{we set 200 as the maximum number of iterations.} 
%
\reedit{At convergence, the proposed approach outperforms the FD-OMA scheme, while it is very close to the results obtained by the other FD-NOMA algorithms.}
\begin{figure}[tb]
\begin{center}
\input{tikz/wmmse_rate_iteration_opt_weight_2}
\caption{$U$ vs number of iterations, $\alpha_w\hspace{-3pt}=$2, for different number of users $N + M$ (blue for 10, red for 30, magenta for 50).}
\vspace{-0.2cm}
\label{fig:convergence}
\end{center}
\end{figure}


\reedit{ Table~\ref{TableComplexity}, shows the complexity of the various schemes, computed considering a) the term that dominates the number of elementary operations  and b) the number of expected iterations.
The lightest scheme is the FD-OMA, which however has by far the worst performance. Among the best-performing schemes, the WMMSE algorithm is the one with the best complexity.}

\begin{table}
\centering
\scriptsize
\begin{tabular}{crr
} 
\toprule
Algorithm & Complexity per iteration & Number of iterations\\
\midrule
REF & $O( F M^2 N^2 )$ & $\approx500$\\
SCA & $O( F M^2 N^2 )$ & $\approx50$\\
WMMSE & $O(2 F (M + N))$ & $\le200$\\
FD-OMA & $O(F (M + N))$ & $\le 100$\\
\bottomrule
\end{tabular}
\caption{Complexity comparison of the algorithms.} 
\vspace{-0.5cm}
\label{TableComplexity}
\end{table}

\begin{figure}[b]
\centering
%
%
\begin{tikzpicture}

\begin{semilogyaxis}[%
width=6.5cm,
height=4cm,
xmin=0,
xmax=200,
xlabel={iterations},
ymin=1e-2,
ymax=50,
ylabel={$R$ [bps/Hz]},
xmajorgrids,
xminorgrids,
ymajorgrids,
legend style={at={(0.734,0.01)}, anchor=south east}
]
\addplot [color=black]
  table[row sep=crcr]{%
1	1.01489042707874\\
2	1.44618677128716\\
3	2.99631659862268\\
4	8.51539114389914\\
5	10.7384720357049\\
12	10.9140837184305\\
18	11.0779146947506\\
24	11.2566760391474\\
29	11.4193004645914\\
34	11.596719094653\\
38	11.7511114767421\\
42	11.9183126653962\\
46	12.1001125049064\\
50	12.2983687713832\\
54	12.5147261600844\\
58	12.7499848807207\\
62	13.0028318321393\\
73	13.7169552969284\\
75	13.8286794678255\\
77	13.9278359433423\\
79	14.012678534181\\
81	14.0826444595606\\
83	14.1383545865176\\
85	14.1813402072594\\
88	14.2264433164832\\
91	14.2545468367968\\
95	14.2755314801136\\
101	14.2891540257966\\
113	14.2964158567222\\
185	14.307341244547\\
200	14.3094745339667\\
};
\addlegendentry{$R_{1,3,f}$}

\addplot [color=black, dashed]
  table[row sep=crcr]{%
1	1.01792161194723\\
2	1.44633267595779\\
3	2.99682355438028\\
4	8.53952485242277\\
5	10.8546365912183\\
11	11.0165930284798\\
17	11.1933186319738\\
22	11.3543042892251\\
27	11.5304163522273\\
31	11.6843400055524\\
35	11.8520878916883\\
39	12.0362714677618\\
42	12.1871429192646\\
45	12.3507942930034\\
48	12.5292645077116\\
51	12.724991841718\\
54	12.9408709783339\\
56	13.0976342304244\\
58	13.2659562381774\\
60	13.4470075800413\\
62	13.6419648022416\\
64	13.8519141810237\\
66	14.0776934177264\\
68	14.3196503642016\\
70	14.5773053893446\\
72	14.8489295613303\\
75	15.2745309145333\\
78	15.7033851243519\\
80	15.9772042345345\\
82	16.230717281683\\
83	16.3473105136244\\
84	16.4560653476292\\
85	16.5564358633676\\
86	16.6481009930088\\
87	16.7309656266303\\
88	16.8051462312212\\
89	16.8709437669035\\
91	16.9792978257354\\
93	17.0607019564137\\
95	17.1204139402028\\
97	17.1634229495356\\
99	17.1939938538626\\
102	17.2237430790624\\
106	17.2451609341811\\
112	17.2587103303346\\
124	17.265855399694\\
198	17.2770382693612\\
200	17.2773225520856\\
};
\addlegendentry{$R_{1,f}$}

\addplot [color=red]
  table[row sep=crcr]{%
1	0.920894695957458\\
2	0.657500228487123\\
3	0.192102391468353\\
4	0.00309757143301681\\
7	0\\
200	0\\
};
\addlegendentry{$R_{2,3,f}$}

\addplot [color=red, dashed]
  table[row sep=crcr]{%
1	0.980420874646882\\
2	0.658131934128733\\
3	0.192111377770942\\
4	0.00309769367086687\\
7	0\\
200	0\\
};
\addlegendentry{$R_{2,f}$}
\end{semilogyaxis}
\end{tikzpicture}%
\caption{Rates  $R_{k,f}$ (solid lines) and $R_{k,s,f}$ (dashed lines) vs  number of iterations for a specific simulation instance.}
\vspace{-0.5cm}
\label{fig:wMMSE:SIC}
\end{figure}

Fig.~\ref{fig:wMMSE:SIC} refers to an instance of the simulation scenario with $M = N = 3$. The strong user is $s=3$ and Fig.~\ref{fig:wMMSE:SIC} plots as straight lines the rates $R_{k,f}$ of all downlink weak users $k\in\mc{D}^w_f=\{1,2\}$  and as dashed lines the rates $R_{k,s,f}$ of the same users measured at the downlink strong user $s=3$  for a given channel $f$ vs the iteration number. Consistently with Theorem~\ref{theo:twouser}, the number of active downlink weak users \edit{on channel $f$} converges to one\edit{, compliant with the exclusivity of resource assignment}.  Moreover, the dashed lines are clearly above the straight lines, thus fulfilling constraint~\eqref{WMMSE:WSWW}. Results collected for any other instances show the same behaviour. 




Fig.~\ref{fig:fairness} shows the impact of different values of $\alpha_w$ vs the number of users $N+M$ by plotting the Jain's fairness index (\edit{solid line}, left y-axis) and the spectral efficiency $R / F$ (\edit{dashed line}, right y-axis). The rate considered here is not weighted, i.e., $R = \sum_{i,f} R_{i,f}(\mb{P},x)$, showing the effect of the increased fairness between strong and weak users in terms of the achievable throughput. As expected, a higher $\alpha_w$ leads to higher fairness and lower spectral efficiency. For $N+M>40$, the fairness of the system does not depend on $\alpha_w$.

\begin{figure}[th]
\centering
%
%
\definecolor{mycolor1}{rgb}{0.00000,0.44700,0.74100}%
\definecolor{mycolor2}{rgb}{0.85000,0.32500,0.09800}%
\definecolor{mycolor3}{rgb}{0.92900,0.69400,0.12500}%
\definecolor{mycolor4}{rgb}{0.49400,0.18400,0.55600}%
\begin{tikzpicture}

\pgfplotsset{set layers, width=6.2cm, height=4.5cm}

\begin{axis}[%
width=5.5cm,
height=3.2cm,
scale only axis,
axis y line*=left,
xmin=10, xmax=50,
xlabel={$N+M$},
ymin=0,
ymax=0.6,
ylabel={Jain's Fairness index},
xmajorgrids,
ymajorgrids,
legend style={at={(1,0.5)}, anchor=east},
]
\addlegendimage{only marks, color=blue, mark=*, mark options={solid, fill=white}}
\addlegendentry{$\alpha_w = 2$}
\addlegendimage{only marks, color=red, mark=triangle*, mark options={solid, fill=white}}
\addlegendentry{$\alpha_w = 4$}
\addlegendimage{only marks, color=mycolor3, mark=diamond*, mark options={solid, fill=white}}
\addlegendentry{$\alpha_w = 6$}

\addplot [color=blue, mark=*, mark options={solid, fill=white}, forget plot]
  table[row sep=crcr]{%
10	0.33289803166612\\
20	0.281142268242227\\
30	0.195568202544003\\
40	0.146398551455924\\
50	0.130782456486665\\
};

\addplot [color=red, mark=triangle*, mark options={solid, fill=white}, forget plot]
  table[row sep=crcr]{%
10	0.398575620206478\\
20	0.327549221059283\\
30	0.199617860327671\\
40	0.148571648985736\\
50	0.130109262096158\\
};

\addplot [color=mycolor3, mark=diamond*, mark options={solid, fill=white}, forget plot]
  table[row sep=crcr]{%
10	0.502230603906575\\
20	0.379259441075757\\
30	0.223196360257509\\
40	0.162878459310272\\
50	0.130239058643066\\
};



\end{axis}

\begin{axis}[%
width=5.5cm,
height=3.2cm,
scale only axis,
axis x line*=none,
axis y line*=right,
hide x axis,
xmin=10, xmax=50,
ymin=39, ymax=62,
ylabel={Sp. eff. $R / F$ [bps/Hz]}, 
yticklabel pos=right,
ytick={36, 42, 48, 54, 60},
yticklabels={6, 7, 8, 9, 10},
]
\addplot [color=blue, dashed, forget plot, mark=*, mark options={solid, fill=white}, forget plot]
  table[row sep=crcr]{%
10	47.353249926435\\
20	51.6437146711475\\
30	55.366242730561\\
40	57.8143259299786\\
50	60.4360714866562\\
};
\addplot [color=red, dashed, forget plot, mark=triangle*, mark options={solid, fill=white}, forget plot]
  table[row sep=crcr]{%
10	44.733692906074\\
20	48.7832874204941\\
30	53.0750169754424\\
40	55.9658319486214\\
50	58.4482158429739\\
};
\addplot [color=mycolor3, dashed, mark=diamond*, mark options={solid, fill=white}, forget plot]
  table[row sep=crcr]{%
10	43.2488043754602\\
20	48.0890853768619\\
30	52.7553935562827\\
40	55.2606014230539\\
50	58.0015069822764\\
};


\end{axis}
\end{tikzpicture}%
\caption{Jain's fairness (solid lines) and spectral efficiency (dashed lines) vs $N + M$, for different values of $\alpha_w$.}
\vspace{-0.5cm}
\label{fig:fairness}
\end{figure} 

\section{Conclusions \edit{and Future Work}}
In this paper, power and channel allocation problem for multi-carrier non-orthogonal multiple access full duplex systems has been investigated. We have proposed a solution based on the minimization of the weighted mean square error, which benefits of the insights obtained by the problem decomposition. The proposed approach  achieves performance close to the optimum at a fraction of the complexity. 
\edit{Future work  will be focused on how to extend the proposed algorithm to a FD-NOMA MIMO setting, enabling the allocation of more than two active users per direction.} 

\section*{Appendix}
\addtocounter{thm}{-2}
\begin{thm}
Any local optima for \eqref{Rmax2}  allows \reedit{the allocation of} at most two NOMA users per channel in each direction.
\end{thm}
\begin{proof}
We adopt a \emph{reductio ad absurdum} argument and assume that~\eqref{Rmax2} yields a solution where the rate on a given downlink channel, say $f$, is maximized by more than two users. The proof for the uplink direction is similar and is omitted for lack of space. By construction, for each channel there is at most one strong user, so we assume that at the equilibrium there are  one strong user $s$ and two weak users $w_{1}$ and $w_{2}$.  To simplify the notation, we omit the channel index and indicate with $\tilde{R}_{i}$ and $\tilde{P}_{i}$ the rate and the power of user $i$ on channel $f_{0}$. We indicate with 
$\tilde{\sigma}^{2}_{i}=(\sum_{j \in  \mathcal{U}} |h_{j,i}(f)|^2 P_{j,f} + \sigma^2) / |h_{i,i}(f)|^2$ the normalized  noise-plus-(uplink-)interference  for user $i$ and with $\tilde{P}_{0}=\tilde{P}_{s}+\tilde{P}_{w_{1}}+\tilde{P}_{w_{2}}$ the total power transmitted on downlink channel $f$. Consistently with these definitions, it is
$\tilde{R}_{s}= \log_{2}\big(1+\frac{\tilde{P}_{s}}{\tilde{\sigma}^{2}_{s}}\big)$, 
$\tilde{R}_{w_{1}}= \log_{2}\big(1+\frac{\tilde{P}_{w_{1}}}{\tilde{P}_{s}+\tilde{P}_{w_{2}}+\tilde{\sigma}^{2}_{w_{1}}}\big)$ and $\tilde{R}_{w_{2}}= \log_{2}\big(1+\frac{\tilde{P}_{w_{2}}}{\tilde{P}_{s}+\tilde{P}_{w_{1}}+\tilde{\sigma}^{2}_{w_{2}}}\big)$. To prove our theorem we make the non-restrictive assumption that $\tilde{\sigma}^{2}_{w_{2}}>\tilde{\sigma}^{2}_{w_{1}}$ and we need to show that the solution with the three users $s,w_{1},w_{2}$ can not be an equilibrium point  because,  freezing all other parameters of the system,  we can find a solution that yields  a higher weighted rate, by allocating only users $s$ and $w_{1}$. 
In particular, we show that it is more beneficial to allocate all the power $\tilde{P}_{0}-\tilde{P}_{s}=\tilde{P}_{w_{1}}+\tilde{P}_{w_{2}}$ to user $w_{1}$, so that its rate   is $\log_{2}\big(1+\frac{\tilde{P}_{0}-\tilde{P}_{s}}{\tilde{P}_{s}+\tilde{\sigma}^{2}_{w_{1}}}\big)$, rather than sharing it between $w_{1}$ and $w_{2}$.  Accordingly, the inequality to prove is
\begin{equation*} 
\alpha_s\tilde{R}_{s}+\alpha_w \big(\tilde{R}_{w_{1}}+\tilde{R}_{w_{2}}\big)<\alpha_s\tilde{R}_{s}+ \alpha_w\log_{2}\bigg(1+\frac{\tilde{P}_{0}-\tilde{P}_{s}}{\tilde{P}_{s}+\tilde{\sigma}^{2}_{w_{1}}}\bigg),
\end{equation*}
which, after few manipulations, becomes 
%
\begin{equation*} \label{eq: app1}
\frac{\tilde{P}_{0}+\tilde{\sigma}^{2}_{w_{1}}}{\tilde{P}_{s}+\tilde{P}_{w_{2}}+\tilde{\sigma}^{2}_{w_{1}}}\bigg(1+\frac{\tilde{P}_{w_{2}}}{\tilde{P}_{s}+\tilde{P}_{w_{1}}+\tilde{\sigma}^{2}_{w_{2}}}\bigg)<\frac{\tilde{P}_{0}+\tilde{\sigma}^{2}_{w_{1}}}{\tilde{P}_{s}+\tilde{\sigma}^{2}_{w_{1}}},
\end{equation*}
which, in turn, is equivalent to
%
\begin{equation*} 
\label{eq: app2}
1+\frac{\tilde{P}_{w_{2}}}{\tilde{P}_{s}+\tilde{P}_{w_{1}}+\tilde{\sigma}^{2}_{w_{2}}}<\frac{\tilde{P}_{s}+\tilde{P}_{w_{2}}+\tilde{\sigma}^{2}_{w_{1}}}{\tilde{P}_{s}+\tilde{\sigma}^{2}_{w_{1}}}=1+\frac{\tilde{P}_{w_{2}}}{\tilde{P}_{s}+\tilde{\sigma}^{2}_{w_{1}}},
\end{equation*}
which  is always true because $\tilde{\sigma}^{2}_{w_{1}}<\tilde{P}_{w_{1}}+\tilde{\sigma}^{2}_{w_{2}}$.
The solution with three users on the downlink channel $f$ can not be an equilibrium point for~\eqref{Rmax2} and the hypothesis is false.
\end{proof}

\bibliographystyle{IEEEtran}
\bibliography{FDNOMA}

@article{Nhat2021,
	Author = {Q. N. Le and others},
	Journal = {{IEEE} Trans. Green Comm. Net.},
	Number = {1},
	Title = {Full-Duplex Non-Orthogonal Multiple Access Cooperative Overlay Spectrum-Sharing Networks With {SWIPT}},
	Volume = {5},
	Year = {2021}}

@article{Sun2017,
	Author = {Y. Sun and others},
	Journal = {{IEEE} Trans. Commun.},
	Number = {3},
	Title = {Optimal Joint Power and Subcarrier Allocation for Full-Duplex Multicarrier Non-Orthogonal Multiple Access Systems},
	Volume = {65},
	Year = {2017}}

@Article{Ding2016,
  author     = {Z. Ding and others},
  title      = {Impact of User Pairing on {5G} Nonorthogonal Multiple-Access Downlink Transmissions},
  journal    = {IEEE Trans. Veh. Technol.},
  year       = {2016},
  volume     = {65},
  number     = {8},
  pages      = {6010-6023},
  month      = aug,
  doi        = {10.1109/TVT.2015.2480766},
}

@Article{Shi11,
  author        = {Qingjiang Shi and others},
  title         = {An Iteratively Weighted {MMSE} Approach to Distributed Sum-Utility Maximization for a {MIMO} Interfering Broadcast Channel},
  journal       = {IEEE Trans. Sign. Proc.},
  year          = {2011},
  volume        = {59},
  number        = {9},
}

@Article{DellaPenda2019,
  author   = {D. Della Penda and others},
  title    = {Distributed Channel Allocation for {D2D}-Enabled {5G} Networks Using Potential Games},
  journal  = {{IEEE} Access},
  year     = {2019},
  volume   = {7},
  pages    = {11195-11208},
  month    = jan,
  issn     = {2169-3536},
  doi      = {10.1109/ACCESS.2019.2891823},
}

@Article{Yu2006,
  author    = {Yu, Wei and Lui, Raymond},
  title     = {Dual methods for nonconvex spectrum optimization of multicarrier systems},
  journal   = {IEEE Trans. Commun.},
  year      = {2006},
  volume    = {54},
  number    = {7},
  publisher = {IEEE},
}

@Article{Nguyen2019a,
  author    = {Nguyen, Hieu  and others},
  title     = {Joint power control and user association for {NOMA}-based full-duplex systems},
  journal   = {IEEE Trans. Comm.},
  year      = {2019},
  volume    = {67},
  number    = {11},
  publisher = {IEEE},
}

@ARTICLE{Ding2018a,
  author={Z. {Ding} and P. {Fan} and H. V. {Poor}},
  journal={IEEE Wireless Commun. Lett.}, 
  title={On the Coexistence Between Full-Duplex and {NOMA}}, 
  year={2018},
  volume={7},
  number={5},
  doi={10.1109/LWC.2018.2811492}}

@Article{Elbayoumi2020,
  author    = {Elbayoumi, Mohammed and others},
  journal   = {IEEE Commun. Surveys \& Tutorials},
  title     = {{NOMA}-Assisted Machine-Type Communications in {UDN}: State-of-the-Art and Challenges},
  year      = {2020},
  publisher = {IEEE},
}

@Article{Mohammadi2019,
  author    = {Mohammadi, Mohammadali and others},
  journal   = {IEEE Comm. Mag.},
  title     = {Full-duplex non-orthogonal multiple access for next generation wireless systems},
  year      = {2019},
  number    = {5},
  volume    = {57},
  publisher = {IEEE},
}

@ARTICLE{Abrardo2020,
  author={A. {Abrardo} and others},
  journal={IEEE Trans. on Wireless Commun.}, 
  title={{Power and subcarrier allocation in 5G NOMA-FD systems}}, 
  year={2020},
  volume={},
  number={},
  pages={},
  doi={10.1109/TWC.2020.3021036}}

@Article{Tseng2001,
  author    = {Tseng, Paul},
  journal   = {Journal of optimization theory and applications},
  title     = {Convergence of a block coordinate descent method for nondifferentiable minimization},
  year      = {2001},
  number    = {3},
  pages     = {475--494},
  volume    = {109},
  publisher = {Springer},
}

@INPROCEEDINGS{abrardo2018optimal,
  author={Abrardo, Andrea and Moretti, Marco}, 
  booktitle={2018 IEEE ICC Workshops}, 
  title={Optimal Power Allocation for {F}ull-{D}uplex Communications over {OFDMA} Cellular Networks}, 
  year={2018},
  volume={},
  number={},
  pages={1-6},
  doi={10.1109/ICCW.2018.8403646}}

@article{arxProof,
  author = {Andrea Abrardo and Marco Moretti and Fabio Saggese},
  title  = {{WMMSE} resource allocation for {NOMA-FD}},
  year   = {2022},
  journal = {arXiv preprint arXiv:1905.06681},
}

@INPROCEEDINGS{Kimy2013,
  author={Kimy, Beomju and others},
  booktitle={IEEE MILCOM 2013}, 
  title={Non-orthogonal Multiple Access in a Downlink Multiuser Beamforming System}, 
  year={2013},
  volume={},
  number={},
  pages={1278-1283},
  doi={10.1109/MILCOM.2013.218}}

@ARTICLE{Choi2015,
  author={J. Choi},
  journal={IEEE Trans. Commun.}, 
  title={{Minimum power multicast beamforming with superposition coding for multiresolution broadcast and application to NOMA systems}}, 
  year={2015},
  volume={63},
  number={3},
  pages={791–800},
}

@ARTICLE{Chen2016,
  author={Z. Chen and others},
  journal={IEEE Trans. Signal Process.}, 
  title={{On the application of quasi-degradation to MISO-NOMA downlink}}, 
  year={2016},
  volume={64},
  number={23},
  pages={6174–6189},
}

@ARTICLE{Chen2017,
  author={Z. Chen and others},
  journal={IEEE Trans. Veh. Technol.}, 
  title={{MED precoding for multiuser MIMO-NOMA downlink transmission}}, 
  year={2017},
  volume={66},
  number={6},
  pages={5501–5505},
}

@ARTICLE{Saggese2020,
  author={F. Saggese and M. Moretti and A. Abrardo},
  journal={IEEE Wireless Communications Letters.}, 
  title={{A Quasi-Optimal Clustering Algorithm for MIMO-NOMA Downlink Systems}}, 
  year={2020},
  volume={9},
  number={2},
  pages={152-156},
}

\end{document}